\documentclass[11pt]{article}

\author{A.C. den Brinker, H. Der Sarkissian, J.H. Wülbern, B. Balmaekers,\\
M. Padalko, J. Sénégas, R. Springorum and C. Possanzini\footnote{A.C. den Brinker, H. Der Sarkissian, B. Balmaekers and M. Padalko are with Philips Research, Eindhoven (NL).  J.H. Wülbern and J. Sénégas are with Philips Research, Hamburg (De).
R. Springorum and C. Possanzini are with MR R\&D, Philips Healthcare, Best (NL).}}
\title{Camera PPG waveforms at the forehead}
\date{}

\usepackage{graphicx}
\usepackage{amsmath}
\usepackage{geometry}
\usepackage{layout}

\graphicspath{{plots/}}

\begin{document}
\maketitle

\section*{Abstract}

In order to obtain insights into the feasibility of replacing ECG-guided triggering in magnetic resonance imaging (MRI) by a system based on video photoplethysmography (PPG), PPG and ECG data were collected from volunteers in an MRI scanner. PPG waveforms obtained using remote camera PPG directed at the forehead are 
studied in qualitative and quantitative sense over a number of volunteers.
The data analysis considers variations in PPG waveforms across volunteers,
modelling of the waveforms in Fourier series, dependencies of waveforms and features 
on the interbeat interval (IBI) and breath-holding, and models for ECG-blind 
estimation of R-peak position. The main findings are that the PPG waveform depends on the volunteer and that its shape changes with IBI and does not depend on breath-holding in the given scenario. Low-order harmonic models provide accurate approximations to the PPG waveform, where for higher IBI the waveform shows more temporal details. 
Accurate predictions (20\,ms std) of the delays between markers in ECG and PPG appear feasible from a single PPG feature. 

\section{Introduction}
 
Photoplethysmography (PPG) is a technique that measures the pulsation of blood volume based on the subtle color changes of the human skin, which led to various applications in the field of physiological and health monitoring reviewed in \cite{ReviewPPGMed}. Originally, it was introduced as a contact-measurement and is used, e.g., in form of the finger tip cardiac sensor in MRI. More recently, a camera-based variant of PPG has been introduced to remotely monitor the heart rate as described in \cite{RemotePPG,WangBook:2021}.

Widespread use of cardiac MRI in standard clinical care is still hindered by complicated planning and workflow. The application of ECG patches adds to this complexity. Their application requires working time of skilled staff, and it adds to the exam time which lowers patient throughput. Patients need to be touched which increases risk of infection and many have to be shaved which lowers patient comfort. Though ECG signal quality is mostly sufficiently robust for cardiac triggering, sometimes electrodes have to be repositioned, optimal positioning depends on the skill of the operator, and occasionally ECG triggering is not possible at all, especially in obese patients.

Technically, ECG signal quality in the MR system is disturbed by several effects: (i) the magneto-hydrodynamic effect, i.e., Lorentz forces on ions in the flowing blood exposed to the stationary magnetic field cause an additional electric field superimposed on the ECG as described in \cite{MHD}; (ii) switched magnetic gradient fields for spatial encoding of the MR signal and (iii) switched control currents to (de-)tune radio frequency coils induce additional voltages that add to the ECG (see \cite{GradientECG}). Hence, the ECG signal measured in the MR bore becomes distorted, which prevents the ECG to be diagnostic and lowers the robustness of ECG triggering especially in high and ultrahigh field MR systems.

Triggering based on markers in a remote PPG signal does not have these disadvantages
and has been proposed as an alternative \cite{TriggerRemotePPG,Wang:2022}.  
Replacement of ECG triggers by a PPG-based trigger mechanism requires detailed 
knowledge of the PPG signal, its characteristics and distribution of properties over the pertinent population. This paper provides first insights into PPG waveforms specifically targeting this application.

The outline of the paper is as follows. We start with describing the data collection.
We continue with describing the data analysis methods with emphasis on personalized prototype construction, modeling and PPG features. The results of the analysis are shown in Section~4, featuring intraperson variations and R-peak predictability. We close with a discussion and conclusions. The paper is supported by further online material.

\section{Methods: data collection}

The data were acquired on volunteers on MR systems located in test bays at the 
Philips Innovation Center, BIU MR in Best (NL). These test-bays are equipped with \emph{3T Ingenia Elition X} systems including a VitalEye product camera, which provides camera-based respiratory-triggered MRI and a view into the bore. This study has been approved by the Internal Committee Biomedical Experiments of Philips Research, and informed consent has been obtained from each volunteer.

Volunteers were set up as for a conventional cardiac exam using the anterior and posterior coil,  ECG-electrodes, and in some volunteers the finger-tip PPU. The cardiac exam included several periods where the participants were asked to hold their breath.

The ECG-signal was manually calibrated after the volunteer was moved into the bore. The software of the commercialized MR system was used to determine the R-peaks of the ECG waveform and the peaks of the contact-PPG waveform. 

The standard mirror was replaced by a system having main mirror and side mirror for improved view of the subjects face.

The available data was collected in two sessions. In the first session, 9 volunteers participated. For a second session 20 volunteers were recruited. One of the volunteers in the second session had an extremely noisy PPG; this data has been excluded from further analysis. For the second session the age range was 30-85\,years, with weight 60-102\,kg and height 162-196\,cm. In total we report on data from 28 participants.

\section{Methods: signal analysis}

\subsection{Prototype creation}

For prototype creation, PPG signal segments created by an automated 
skin selection mechanism were used. Segments with 50\% overlap in time were taken and analyzed. The segments were cut into cardiac cycles. The signal in each a cardiac cycle was
resampled to a uniform grid (e.g., $N_s=100$ samples in a cycle),
and the median was taken over the different cycles where the number of cycles is denoted as $N_c$. The median is called the prototype PPG waveform; next to the median, 
the interquartile range (IQR) was determined to obtain a measure for the variability.  
Two methods for dividing the segments into cardiac cycles were used providing 
waveforms $p_i(k)$, where $i$ gives the method ($i=1,2$) 
and $k$ defines the sample within the cardiac cycle $k=1,\cdots,N_s$.
The two methods are as follows.

\noindent
\textbf{Method 1: Extraction by ECG-based segmentation}\\ 
In this extraction method, the ECG signal was used for 
segmentation. The R-peaks of the ECG signal were identified and the PPG signal was resampled and segmented according to this timing. 

\noindent
\textbf{Method 2: Extraction using PPG signal only}\\ 
In this extraction method, no external timing information was used.  
As marker for time zero, the downgoing zero-crossing was used.
This instance closely corresponds to the steepest slope of the zero-mean PPG signal.

The prototypes of both methods are denoted as $p_1$ and $p_2$, respectively.
Cardiac cycles with a duration much shorter or longer than the median IBI were discarded.
The prototypes $p_1$, $p_2$ are compared in the results section.
To make such a comparsion, the variation in the measurements need to be taken into account. The variation of the waveform in the various cycles was tracked and was quantified by the interquartile range (IQR) at each sample in the cycle. For clarity and simplicity, only the IQR of the $p_1$ prototype is shown in the graphs. In view of the additional processing needed to determine segmentation in the second method, one would expect the second method to provide more noisy results. 

The cardiac cycles were also separated based on whether or not they belonged to a breath-hold epoch. In this way, two PPG prototypes were created for each participant with one reflecting the average PPG waveform during breath-hold periods and the other in other periods. In order to limit the amount of work associated with the required manual annotation, this was done for a limited number of participants.

For further intraperson data analysis, the cardiac cycles were separated into 3 bins depending on the IBI. Per bin a prototype was constructed, yielding three prototype per volunteer corresponding to a low, medium and high IBI.  Systematic changes in features of these prototypes (signal strength and position of the maximum and zero-crossings) were studied.

\subsection{Harmonic model}

The discrete-time prototype was modeled in a Fourier series
 \begin{equation}
   p(k) = \sum_{m=0}^{M} A_m \cos( 2\pi k/Ns+\phi_m) + e(k) 
  \label{eq:HarmonicModel}
 \end{equation}
where $M$ is the order of the harmonic model, $m$ denotes the harmonic component ($m=0$ being the fundamental), $A_m$ the amplitude, $\phi_m$ the phase, and $e(k)$ the non modeled part. In the results section the modeled energy contained in this series is considered as a function of $M$, as well as its dependence on the IBI. 

\subsection{Feature extraction}

\begin{figure}[h]
    \begin{center}
               \includegraphics[width=0.5\textwidth]{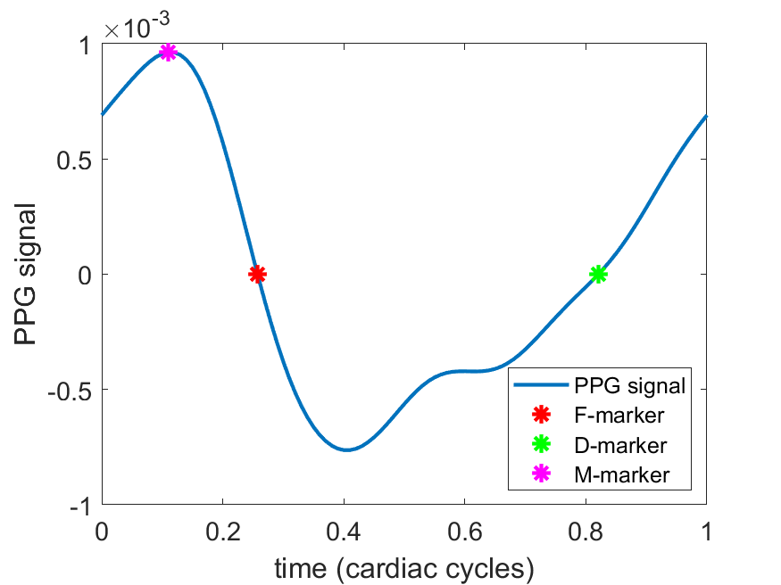}
    \end{center}
    \caption{PPG prototype markers M, F and D.}
    \label{fig:protomarkers}
\end{figure}

From the PPG prototype, a number of feature can be extracted. Although numerous 
features can be defined, we consider only four. All of them are related to specific instances in time. 
The first three are the M, F and D feature being the position of the maximum~M
of the PPG signal, the negative-going zero-crossing F, and the positive-going zero-crossing D.
The zero-crossing F closely corresponds to steepest decrease in PPG signal assumed to be the highest incoming flux of the blood volume or its induced tissue effect. 
D corresponds to a decay of blood flux (at the forehead) assumed to occur in the diastolic phase. All three features are markers from the prototype itself and indicated in Fig.~\ref{fig:protomarkers}. The fourth feature is called Z$_H$ and is the position of the positive-going zero-crossing of the phase of the analytic signal~$a$ associated with $p$. The feature Z$_H$ typically precedes the maximum M.

\section{Results}

\subsection{Prototypes: extraction method}

\begin{figure}[h]
    \begin{center}
        \includegraphics[width=0.49\textwidth]{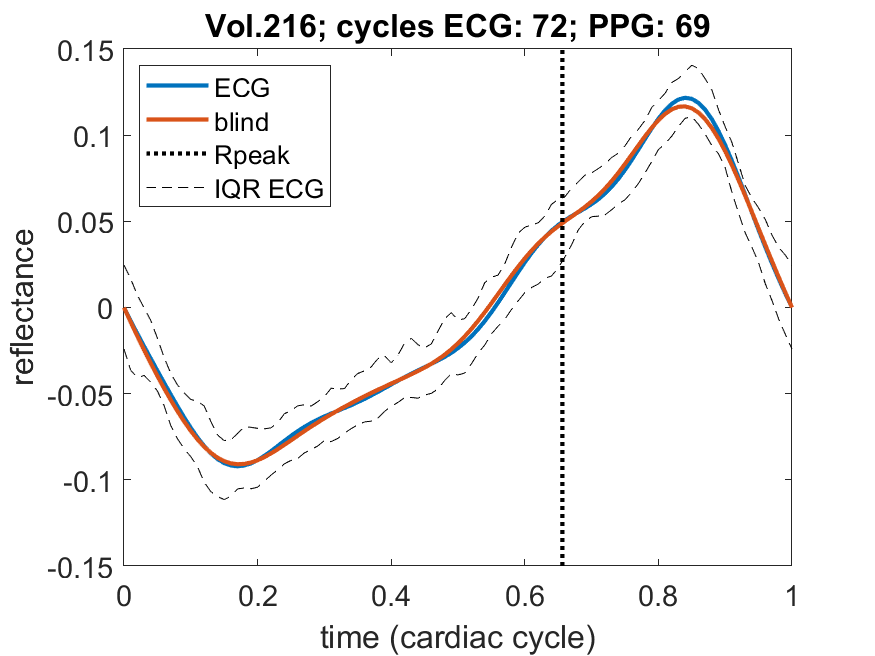}
        \includegraphics[width=0.49\textwidth]{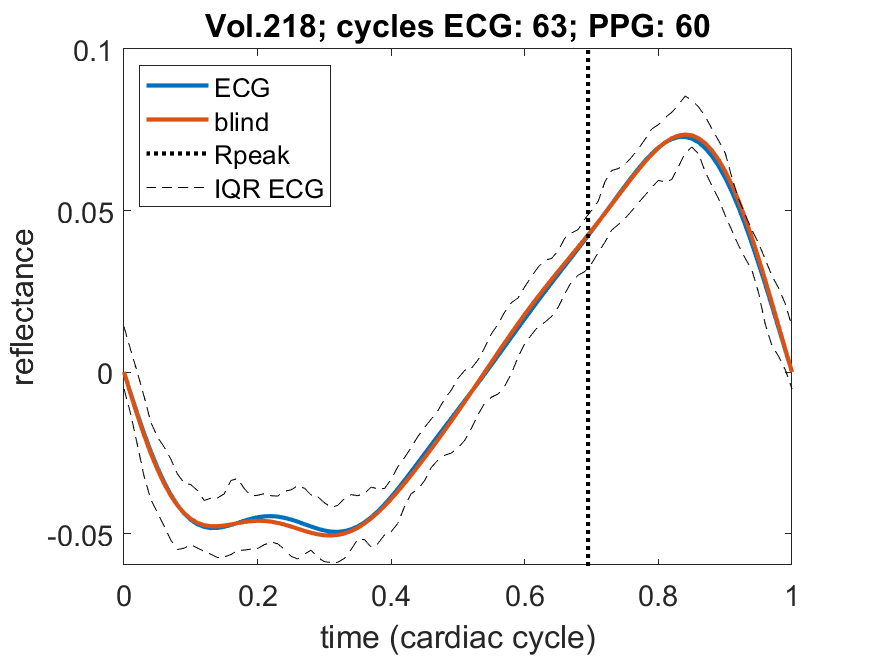}
    \end{center}
    \caption{PPG prototype waveforms of Volunteers 2-16 and 2-18. For both volunteers, two prototypes were constructed where one method is ECG agnostic, and the other is not. The interquartile range (IQR) of the prototype associated ECG-based segmentation is shown as the dashed lines.}
    \label{fig:Waveforms}
\end{figure}

In Fig.~\ref{fig:Waveforms}, the constructed PPG prototypes are shown for two volunteers, where the results from both prototype construction methods are shown. It shows that the ECG-blind method provides prototype waveforms agreeing well with those obtained by using ECG timing information. Apparently, the method of signal segmenting by zero-crossings does not compromise the prototype construction. Qualitatively, the prototypes are biphasic responses, where the positive peak is almost always sharper than the negative one and the downward slope is steeper than the rising one. We also observe that different volunteers provide different waveforms, e.g., the distances between zero-crossings and other waveform details are volunteer dependent. 

One of the participants showed clearly deviant PPG behavior. The common PPG prototype has a relatively narrow peak shortly after the R-peak maximum. This can be understood from the fact that the R-peak is the onset of the heart muscle contraction, which is followed by opening of the valves and a fresh blood bolus being pushed through the vessels, finally to arrive at the observation site where the PPG signal reflects the blood volume change. Deviant behavior was defined as a PPG were the maximum of the prototype is relatively broad (in comparison to the valley) and the maximum occurs at or before the R-peak. It is assumed that some signal disturbance caused this atypical behavior. The data from this participant have been excluded from the next analyses.   

\subsection{Prototypes: breath-holding}

\begin{figure}[h]
    \begin{center}
        \includegraphics[width=0.49\textwidth]{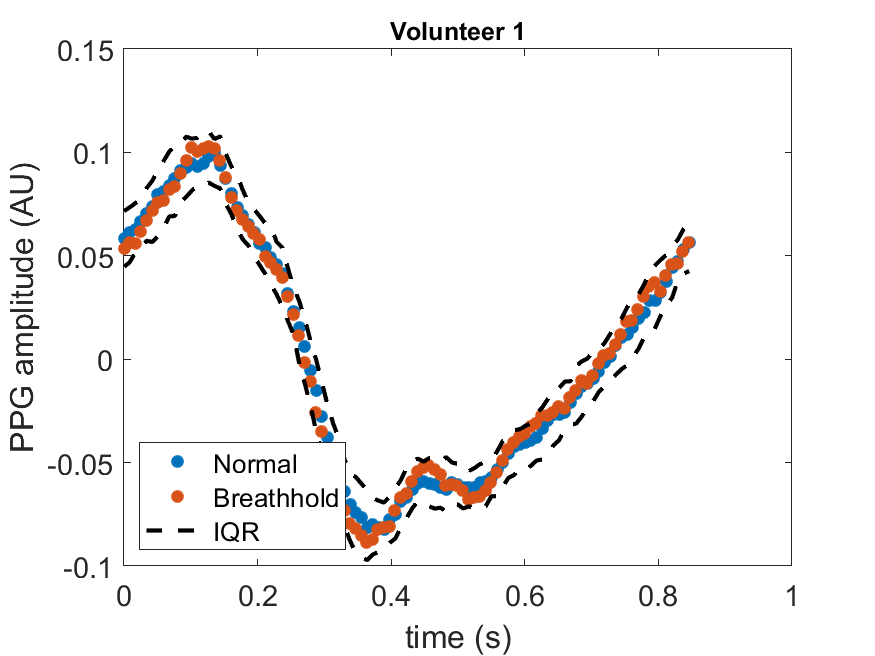}
        \includegraphics[width=0.49\textwidth]{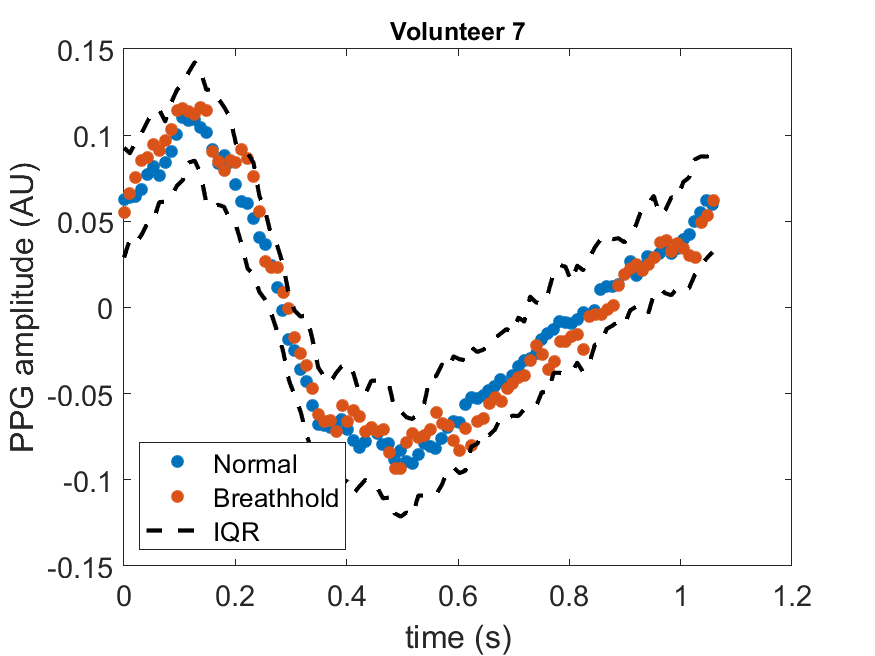}
    \end{center}
    \caption{PPG prototype waveforms of Volunteers 2-01 and 2-07. For both volunteers, two prototypes were constructed for the breath-holding and non-breathholding (normal) epochs. The interquartile range associated with the normal prototype is indicated by the dashed lines.}
    \label{fig:Breathhold}
\end{figure}

For 6 participants of the second session, prototypes were created for the breath-holding epochs and non-breath-holding epochs. The latter epochs thus include preparation and recovery of breath-holding phases. In Fig.~\ref{fig:Breathhold}, the constructed PPG prototypes are shown for two volunteers, where the prototypes for the two conditions are given. The constructed prototypes have not been smoothed nor normalized (in time or amplitude). One PPG prototype is cleaner and of shorter duration, the other longer and more noisy (as inferred from the IQR). For all considered participants, the prototype waveforms from these different epochs agree closely similar to those shown in Fig.~\ref{fig:Breathhold}, i.e., with curves falling inside the IQR range. 

\subsection{Prototypes: IBI dependence}

\begin{figure}[h]
    \begin{center}
        \includegraphics[width=0.49\textwidth]{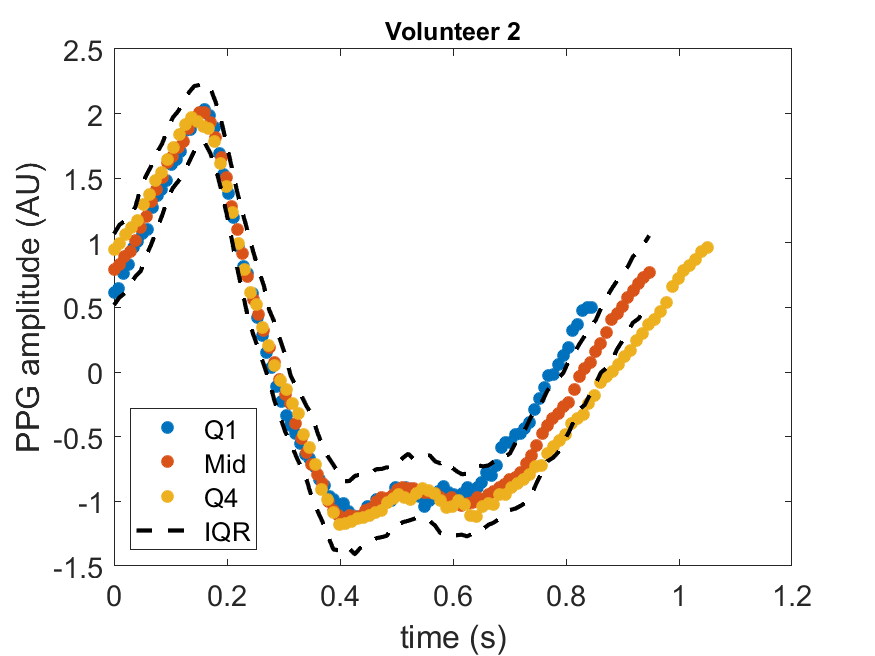}
        \includegraphics[width=0.49\textwidth]{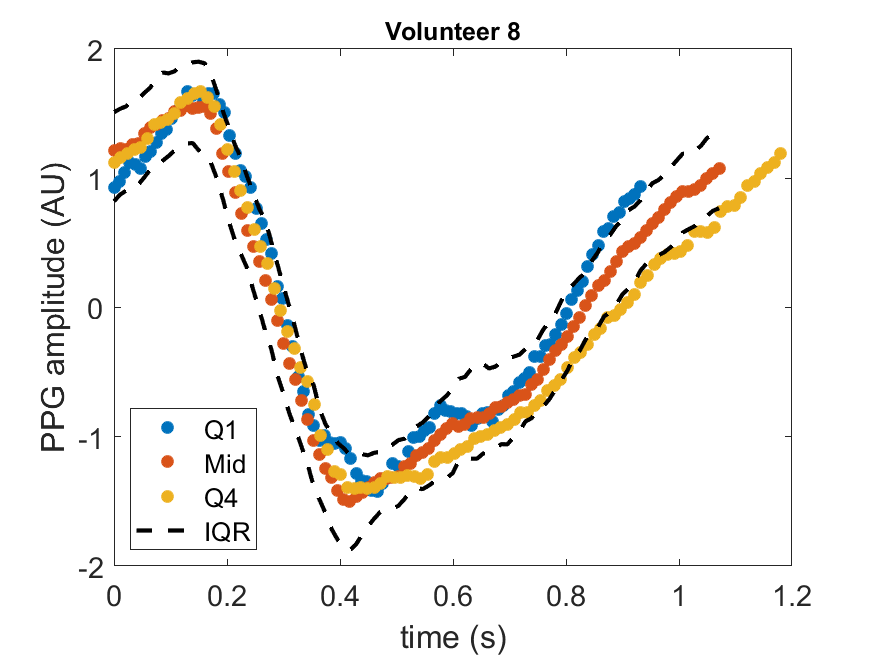}
    \end{center}
    \caption{PPG prototype waveforms of Volunteers 1-06 and 1-08. For both volunteers, three  prototypes were constructed based on IBI subdivision.  The interquartile range of the prototype associated with the central IBI bin is shown as the dashed lines.}
    \label{fig:intraperson}
\end{figure}

For analysis of intraperson variation of the PPG waveforms, the prototypes from ECG-based segmentation were used. Each cardiac cycle was labeled according to its IBI. Three IBI bins were created: the 25\% lowest and highest IBI (denoted as Q1 and Q4) and the remaining mid range (Mid 50\%). 
This was done for each participant separately and for each bin, the corresponding PPG prototype was created. In most cases, the median IBI of both outer bins was less than 5\% from median IBI in the central bin. No clear differences between the prototypes were observed in these cases.
For two volunteers having a large IBI variation within the session, the prototypes are shown in Fig.~\ref{fig:intraperson} where amplitude normalization was applied for clarity of the graphs. These plots suggest that the PPG waveform changes with increasing IBI mainly by time stretching of its rising part; the down-going part appears hardly affected.  

For a more detailed analysis, the prototypes were smoothed by modeling the prototype as a harmonic series consisting of five components: fundamental and four harmonics. The removal of higher harmonic results in smooth prototypes without losing essential information as is shown in the next section. 

\begin{figure}[h]
    \begin{center}
        \includegraphics[width=0.7\textwidth]{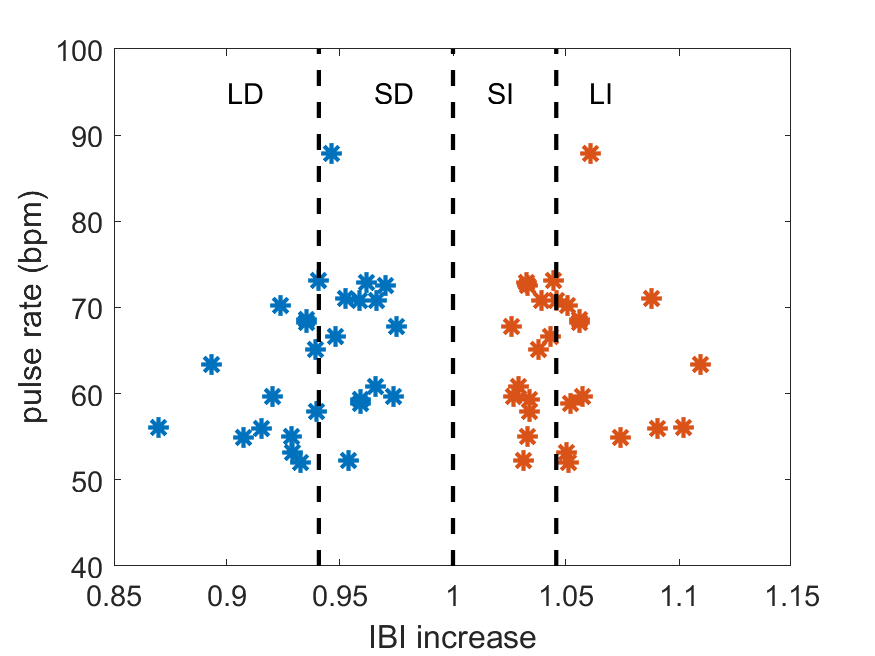}
     \end{center}
    \caption{Intraperson IBI variation categories LD, SD, SI and LI characterized by the
median heart rate of the interquartile range (vertical axis) and the ratio of the outer quartiles and interquartile median IBIs (horizontal axis). Blue asterisks: $R_D$, red asterisks: $R_I$. 
}
    \label{fig:Categories}
\end{figure}

From the smoothed prototypes, four features were extracted. These are the positions of the markers M, F and D (relative to the R-peak) and the signal amplitude at the maximum. These were established for all three bins, with the bins denoted as D, N and I reflecting the decreased, normal and increased IBI categories. Each participant has a median IBI in these three categories denoted as $T_D(p)$, $T_N(p)$ and $T_I(p)$ with $p$ indicating the participant. Considered were cases with small (S) and large (L) increases and decreases in duration quantified by the ratios
 \begin{equation}
  R_D(p) = T_D(p)/T_N(p) \mbox{ and } R_I(p) = T_I(p)/T_N(p). 
 \end{equation}
The D data was split into two sets: participants with $R_D$ less than the median or not, splitting the D data in 13 and 14 participants, respectively. The sets are denote as LD and SD corresponding to participants exhibiting a large and small decrease of IBI. Similarly, the  I data was split into two sets: participants with $R_I$ greater than the median or not, splitting the I data in 13 and 14 participants, respectively. The sets are called LI and SI indicated a large and small increase of IBI. A scatter plot of the ratios $R_I$ and $R_D$ as a function of pulse rate (derived from the N data) is shown in Fig.~\ref{fig:Categories} and clarifies the subdivision into the four categories.

\begin{figure}[h]
    \begin{center}
        \includegraphics[width=0.49\textwidth]{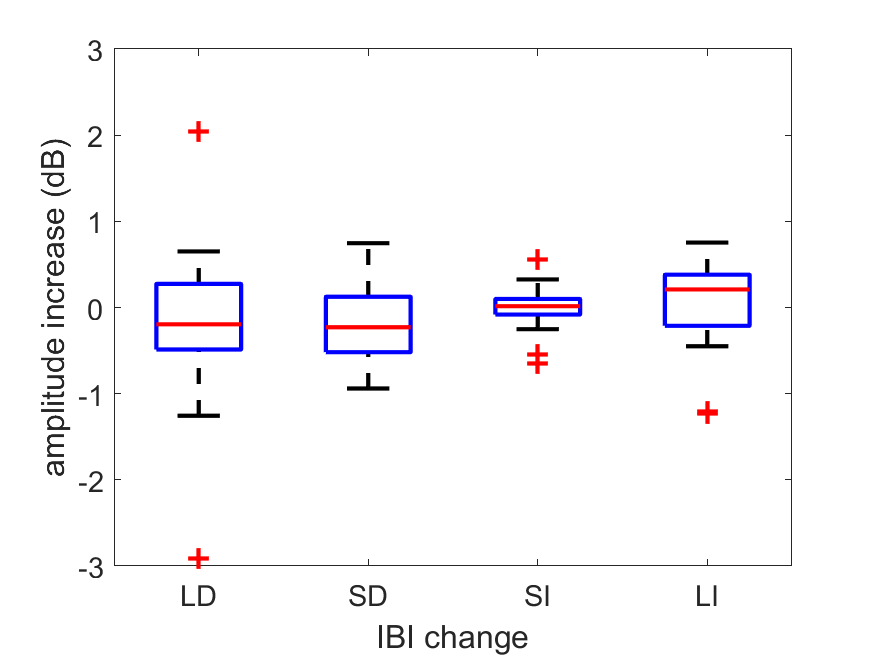}
	\includegraphics[width=0.49\textwidth]{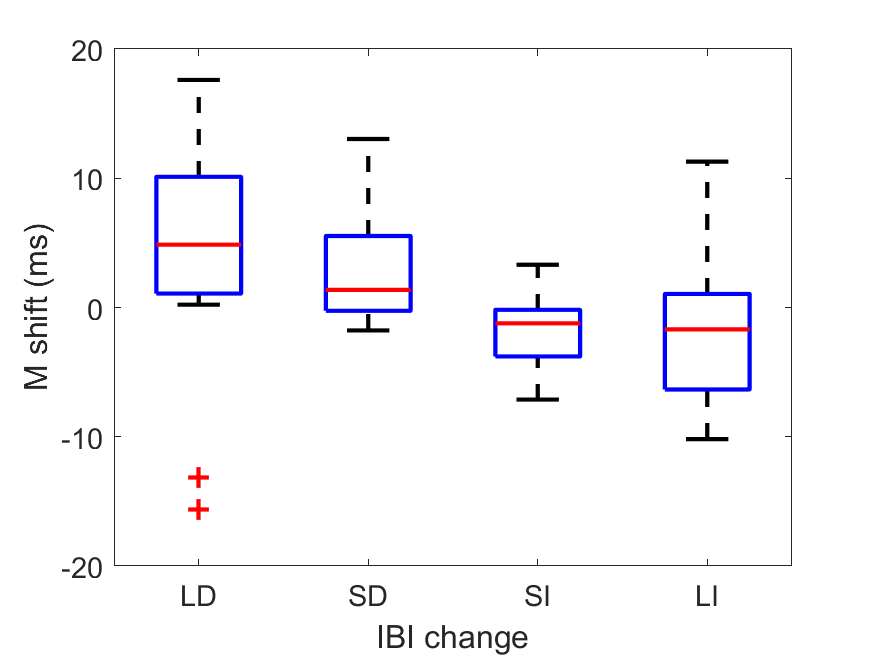}

\includegraphics[width=0.49\textwidth]{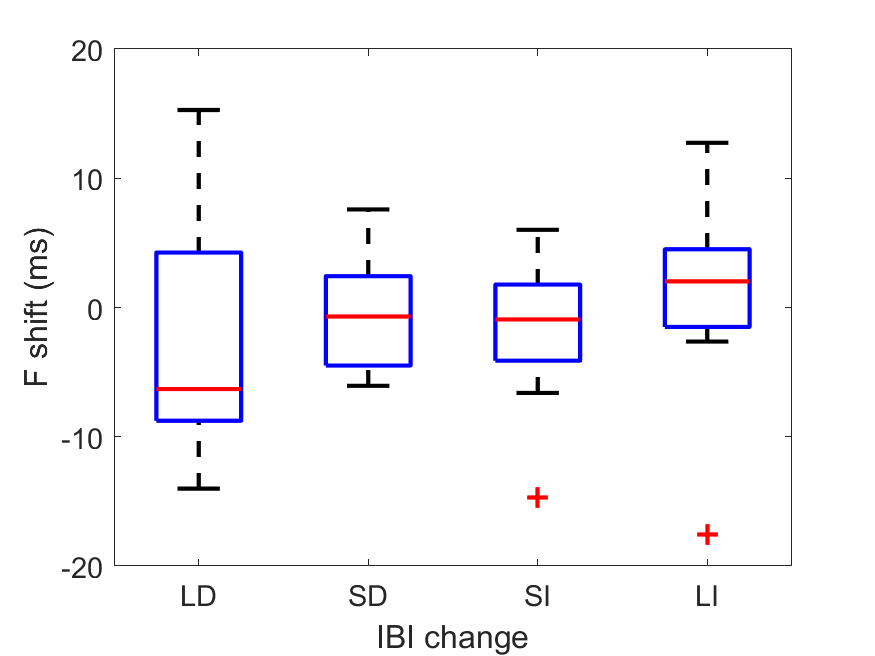}
\includegraphics[width=0.49\textwidth]{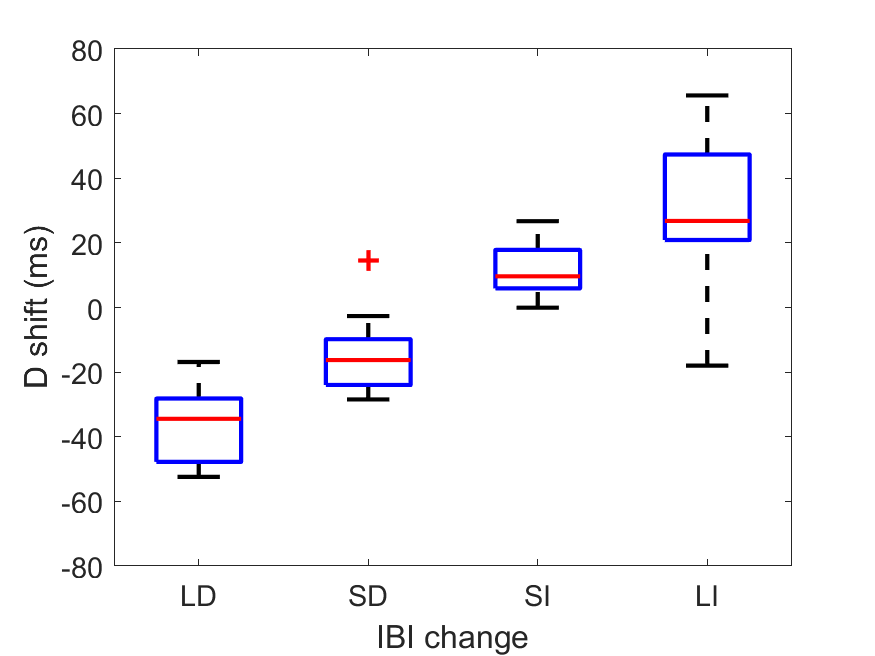}

    \end{center}
    \caption{Boxplots of feature variation in the four intraperson IBI variation categories
LD, SD, SI and LI (see text). Top row: amplitude change and position shift of the maximum. 
Bottom row: position shift of F and D marker.}
    \label{fig:FeatureChange-IBI}
\end{figure}

Boxplots of the amplitude and the position change associated with four categories of  IBI change are given in Fig.~\ref{fig:FeatureChange-IBI}. The plots indicate the RD distance changes significantly and consistently with the IBI change. For the other metrics (amplitude, RM and RF distance) it would require more data to support the hypothesis of a change, and if so, the change would be small (amplitude change of less than 1\,dB, position shifts of several ms).

\clearpage

\subsection{Harmonic model}

\begin{figure}[h]
    \begin{center}
               \includegraphics[width=0.8\textwidth]{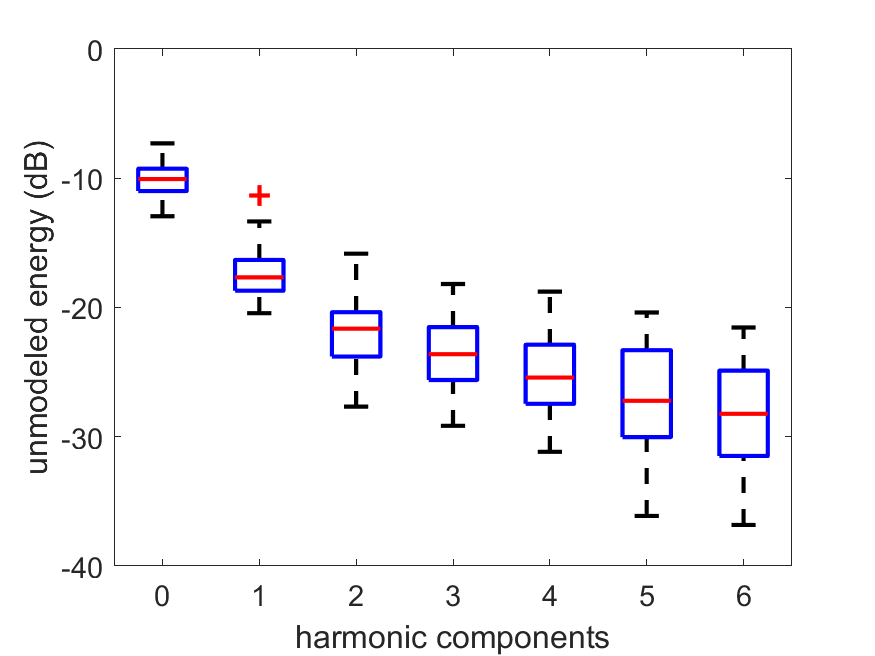}
    \end{center}
    \caption{Boxplot of unmodeled energy in the harmonic model for PPG prototype waveforms as a function of the number of harmonics.}
    \label{fig:EnergyOrder}
\end{figure}

The prototype PPG waveforms were modeled in a Fourier series of different order $M$, see (\ref{eq:HarmonicModel}). The fraction of unmodeled energy (in dB) is shown as a function of order of the harmonic series over all 25 volunteers as a boxplot in Fig.~\ref{fig:EnergyOrder}. Beyond order $M=2$, central 50\% ranges of adjacent orders overlap to a high degree. It shows that with $M=2$ almost all energy is captured by the truncated Fourier series: for 50\% of the cases the modeling error is less than 20\,dB.

\begin{figure}[h]
    \begin{center}
               \includegraphics[width=0.8\textwidth]{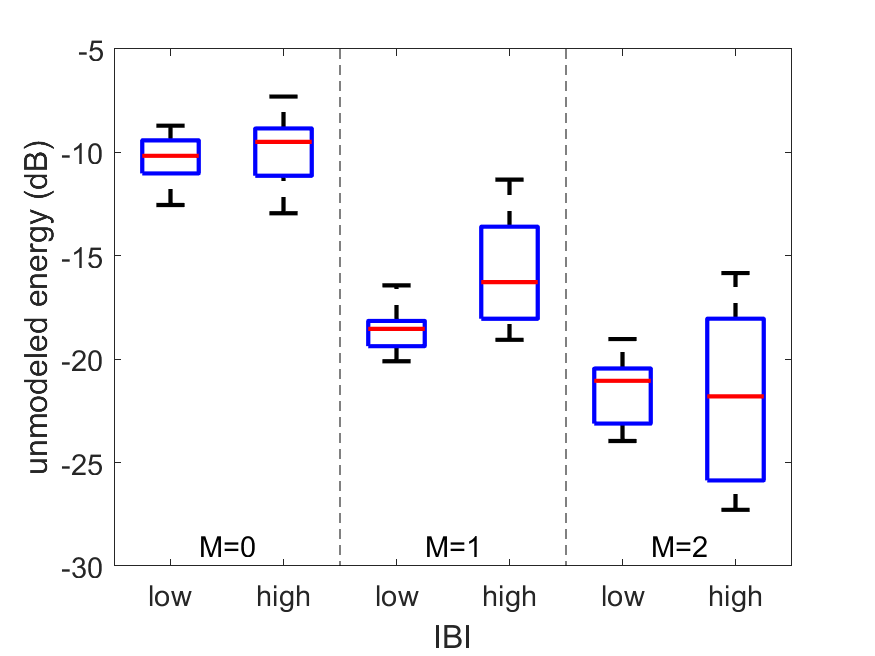}
    \end{center}
    \caption{Boxplot of unmodeled energy in the harmonic model ($M =0,1,2$) for PPG prototype waveforms associated with low and high IBI.}
    \label{fig:EnergyIBI}
\end{figure}

The modeling was also analyzed for dependence on IBI. This is done by creating 3 IBI bins and comparing the results of two outer bins, i.e., containing the 9 lowest and highest IBIs. In Fig.~\ref{fig:EnergyIBI}, the unmodeled energy of this subdivision is shown suggesting that for low IBI (high heart rate) the modeling with 2 components gives better results than in case of high IBI. For orders $M=0$ and $2$, the distributions overlap. In other words, the first harmonic component is stronger for high IBI than for low.

\clearpage
\subsection{PPG features}

\begin{figure}[h]
    \begin{center}
               \includegraphics[width=0.8\textwidth]{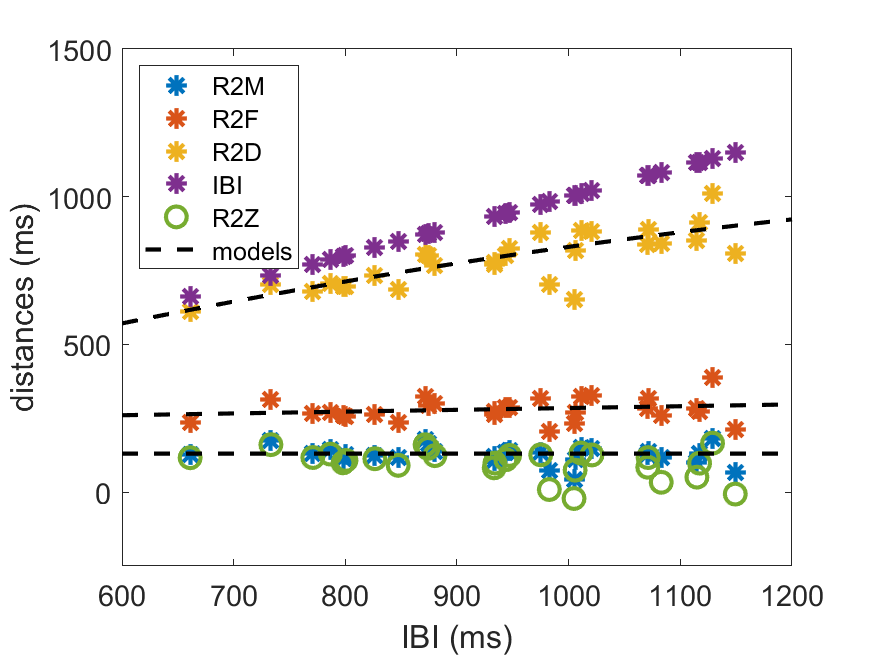}
    \end{center}
    \caption{Scatter plots of distance of PPG markers to the R-peak as determined from prototypes.}
    \label{fig:Markers2Rpeak}
\end{figure}

In this section the PPG markers and their relation to the ECG timing is considered.
Fig.~\ref{fig:Markers2Rpeak}, the various PPG features (M, F, D and Z$_H$) are plotted as a function of IBI. We see that there is a general trend over the various volunteers.
For the M, F and D features, this can be modeled as a low-order polynomial (dashed lines).
There occur however considerable deviations from these models per volunteer: deviations up to 100\,ms can be observed for individual cases. 

For ECG-blind triggering, it would be convenient if the position of the R-peak can be predicted from the PPG signal itself. Essentially, this means that it is desired that the offset between any of the features and the R-peak can be accurately predicted. This is similar to the attempts of pulse arrival time (PAT) prediction. Various combinations of PPG features were tried. Based on the current data, the most successful prediction $\hat{d}$ was that of the distance~$d$ between the R-peak and M. This distance~$d$ can be modeled as 
 \begin{equation}
   d = \hat{d} + e = K_1 + K_2 d_{ZM} +e
 \end{equation}
where $e$ is the prediction error, $K_1$ and $K_2$ are constants and $d_{ZM}$ is the distance between the zero-crossing of the phase of the analytic signal and the maximum M of the PPG signal, where the distance between the PPG markers Z and M is derived from the ECG-blind prototype construction and the R-peak to Z distance $d$ from the prototype using the ECG segmentation. The experimental cumulative density function of the prediction error~$e$ error for this model and also for the case that the model is just a constant. These curves highlight that using $d_{ZM}$ reduces the outliers: the 90\% confidence interval reduces from 108 to 61\,ms. The error spread is quantified by its standard deviation $\sigma = 20$\,ms. Taking more predictors into account (e.g., distances between M and F or F and D, etc.) did not improve the prediction error significantly.
 
\begin{figure}[h]
    \begin{center}
               \includegraphics[width=0.8\textwidth]{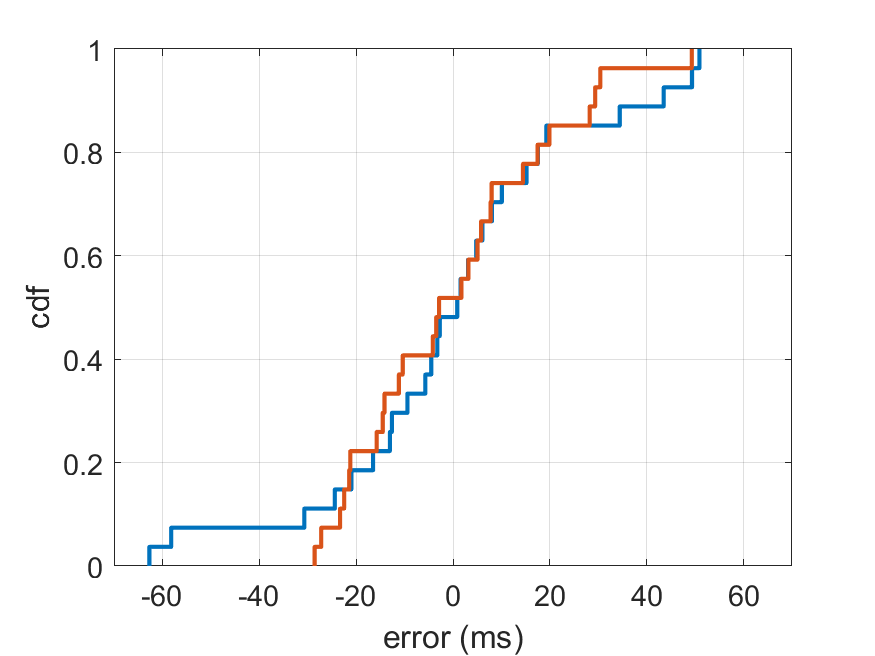}
    \end{center}
    \caption{Experimental cumulative density function of $e$.The blue curve is the  distribution of $d-\mu_d$, with $\mu_d$ the mean value of $d$. The red curve is the error distribution of $e$ with optimized parameters $K_1$ and $K_2$.}
    \label{fig:MarkerPrediction}
\end{figure}
\clearpage

\section{Discussion}


For the development of a PPG-based trigger mechanism for MR, it is encouraging that there is much qualitative agreement between the PPG signals between different subjects. However, the data pool is still limited and deviant behavior may have been missed. The data reveals variation between PPGs from different participants as well as variations per person during the recording period where these intraperson variations included arrythmias for some participants. The recruitment only targeted healthy patients and therefore differentiation of the PPG waveform for cardiac patients could not be considered.

We studied intraperson variability where a subdivision of the data according to IBI was considered. Not addressed were long-term intraperson variation, e.g., variation of the PPG signal of a specific subject at different days. The constructed PPG waveforms suggest that various phases in the PPG vary differently with a change in IBI. The phase of blood volume increase appears to be as good as constant while the phase of blood volume decrease clearly decelerates with increased IBI. This agrees qualitatively with ECG knowledge: there the systolic phase is less affected by a change in IBI than the diastolic phase. The models in \cite{ECGsubdiv0}, \cite{ECGsubdiv1} predict that at an IBI of 500\,ms, the systolic and diastolic duration are about equal (250\,ms), while at an IBI of 1500\,ms this changes to a 450 and 1150\,ms subdivision, i.e., a much larger increase in the diastolic phase than in the systolic one. 

The PPG signal at the forehead can be described well by a low order harmonic model.
For the prototypes, a Fourier series with two harmonics provides already an accurate description of the measurement data. The relative signal strength of the harmonic components appears dependent on the IBI.

A model was created for blind prediction of the R-peak position from the PPG prototype. 
This model was trained only for PPG data from forehead and for a limited number of subjects. Further model validation is recommended. We also note that the model is specific to the (used) camera setting: changing the shutter profile may change the data and thus the parameterization of the model. The model is extremely simple, depending on a single PPG feature only. In contrast, in many blood pressure research, multiple PPG features have been used and tested for blood velocity, PAT and PPT extraction. Similar to our finding,  a single feature was proposed in \cite{Addison:2016} for PTT of contact PGG at different body sites (finger, forehead, ear). This feature (the maximum PPG slope) was tested for our purpose as well but provided poor prediction results.

The PPG signal was determined from a camera operating at frame rate of 40\,fps. This is much lower than the sampling rate of a typical ECG signal, yet it is considered as a sufficiently high rate for accurate triggering for the following reason. Accurate triggering means that the produced images are not suffering from blurring due to trigger jitter.
In cine imaging, the cardiac cycles are binned into phases. The number of phases dictated by image quality: a too low number would lead to blurring the constructed image. A number of $N_p=40$ phases is considered being definitely at the safe side: lower numbers are common in practice ($N_p=10$ to $30$). For a heart rate of 60\,bpm, $N_p=40$ effectively corresponds to a sampling frequency of 40\,Hz. Since the cine sequences do not suffer from motion artifacts, the timing accuracy provided when operating the camera at 40\,fps is presumed sufficient. 

\section{Conclusions}

Application of video-based PPG monitoring for MR triggering requires detailed knowledge of the PPG signal. The PPG signal at the forehead has been studied for inter- and intra-person variability by constructing prototype PPG signals per participant. In qualitative terms, the PPG prototype is a biphasic signal with a relatively fast decrease and a slower PPG rise. Low-order Fourier series allow accurate modeling of the prototype. The prototype is dependent on subject and IBI, but is not dependent on whether or not the participant is executing a breath-hold procedure. Variation of the prototype with IBI is mainly in the rising part of the PPG signal. Person-specific prototype signals with equal accuracy can be constructed with or without timing information from the ECG signal. Simple models allow to predict the R-peak position within the prototype from prototype features with an STD of around 20\,ms for the prediction error.   

\section*{Acknowledgement}

Thanks for support to E. Coezijn, A. Kok, and P. Ramamurthy.
   
\bibliographystyle{unsrt}
\bibliography{MRtiming.bib}

\end{document}